\documentclass[aps,prl,reprint,superscriptaddress,showpacs]{revtex4-1}
\usepackage{graphicx}
\usepackage{amsmath,amssymb}
\usepackage{color}
\usepackage[caption=false]{subfig}
\usepackage{textcomp}

\newcommand{\figwidth}{0.45\textwidth}

\begin{document}


\title{Permeability of porous materials determined from the Euler characteristic}



\author{Christian Scholz}
\affiliation{2.\ Physikalisches Institut, Universit\"at Stuttgart, 70569 Stuttgart, Germany}
\author{Frank Wirner}
\affiliation{2.\ Physikalisches Institut, Universit\"at Stuttgart, 70569 Stuttgart, Germany}
\author{Jan G\"otz}
\affiliation{Lehrstuhl f\"ur Systemsimulation, Friedrich-Alexander Universit\"at Erlangen-N\"urnberg, 91058 Erlangen, Germany}
\author{Ulrich R\"ude}
\affiliation{Lehrstuhl f\"ur Systemsimulation, Friedrich-Alexander Universit\"at Erlangen-N\"urnberg, 91058 Erlangen, Germany}
\author{Gerd E. Schr\"oder-Turk}
\affiliation{Institut f\"ur Theoretische Physik, Friedrich-Alexander Universit\"at Erlangen-N\"urnberg, 91058 Erlangen, Germany}
\author{Klaus Mecke}
\affiliation{Institut f\"ur Theoretische Physik, Friedrich-Alexander Universit\"at Erlangen-N\"urnberg, 91058 Erlangen, Germany}
\author{Clemens Bechinger}
\affiliation{2.\ Physikalisches Institut, Universit\"at Stuttgart, 70569 Stuttgart, Germany}
\affiliation{Max-Planck-Institut f\"ur Intelligente Systeme, Heisenbergstra{\ss}e 3, 70569 Stuttgart, Germany}


\date{\today}

\begin{abstract}
We study the permeability of quasi two-dimensional porous structures of randomly placed overlapping monodisperse circular and elliptical grains. Measurements in microfluidic devices and lattice Boltzmann simulations demonstrate that the permeability is determined by the Euler characteristic of the conducting phase. We obtain an expression for the permeability that is independent of the percolation threshold and shows agreement with experimental and simulated data over a wide range of porosities. Our approach suggests that the permeability explicitly depends on the overlapping probability of grains rather than their shape.
\end{abstract}


\maketitle

\begin{figure}
	\centering
	\includegraphics[width=\figwidth]{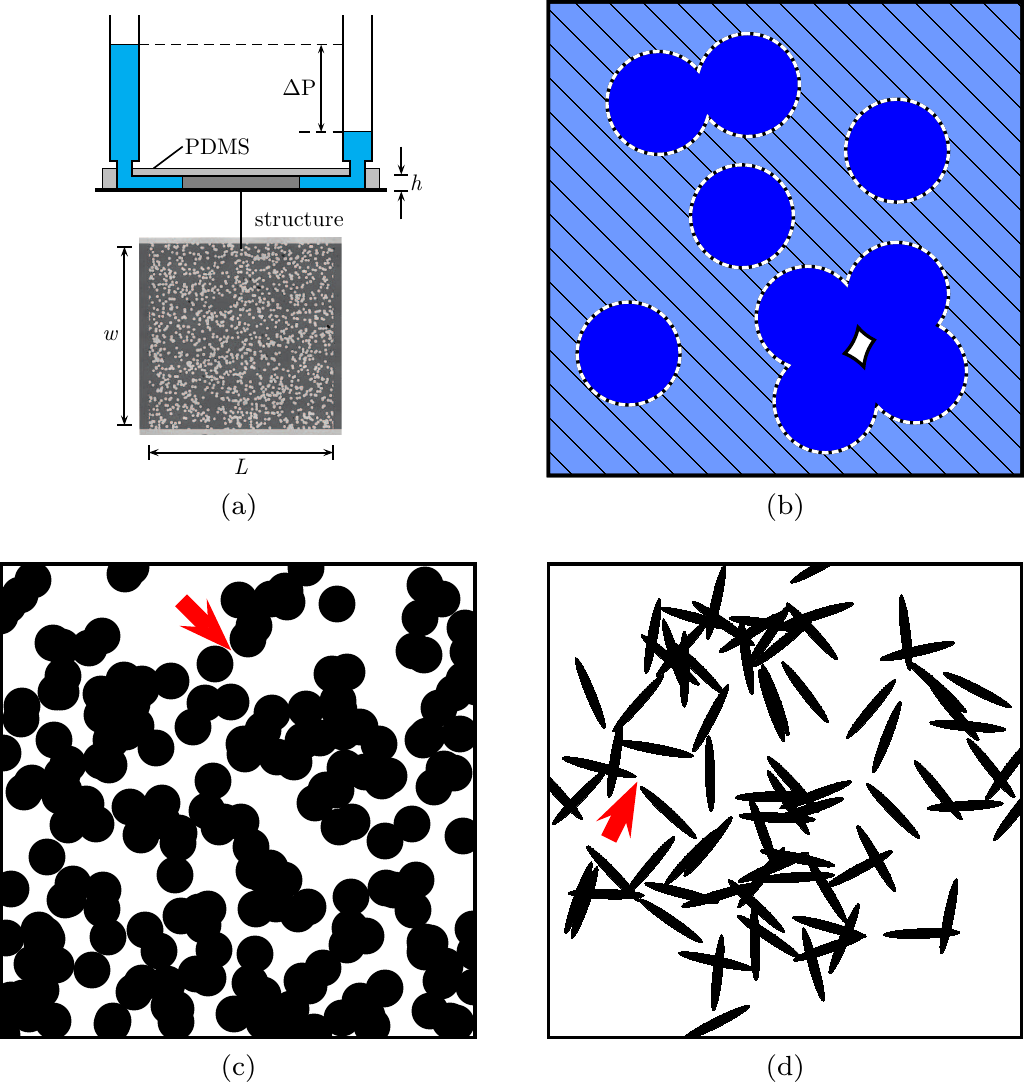}
	\caption{\label{Fig:ExperimentStructures}(Color online) a) Sketch of the experimental setup to measure the permeability through a porous material. b) Illustration of the morphological quantities to describe a porous sample which consists, in this example, of $N=9$ grains (circles) leading to 5 connected solid clusters. The porosity $\phi$ corresponds to the area fraction of the entire liquid phase (hatched + white inclusion) while the open porosity $\phi_o$ is only defined by the conducting phase (hatched). The perimeter is the length of the dashed line and the Euler characteristic is $\chi=-3$ (2 liquid phase components $-$ 5 solid clusters) and $\chi_o=-4$ (the latter neglecting the liquid phase inside the cavity formed in one cluster). Exemplarily we show one realization for the Boolean process for ROMC c) $\phi=0.406$, $\phi_o=0.375$, $N=200$, $\chi_o=-18$ and for ROME d) $\phi=0.805$, $\phi_o=0.78$, $N=80$ and $\chi_o=-17$. The arrows point to the pore throats corresponding to the critical pore diameter $D_c$.}
\end{figure}
\begin{figure}
	\includegraphics[width=\figwidth]{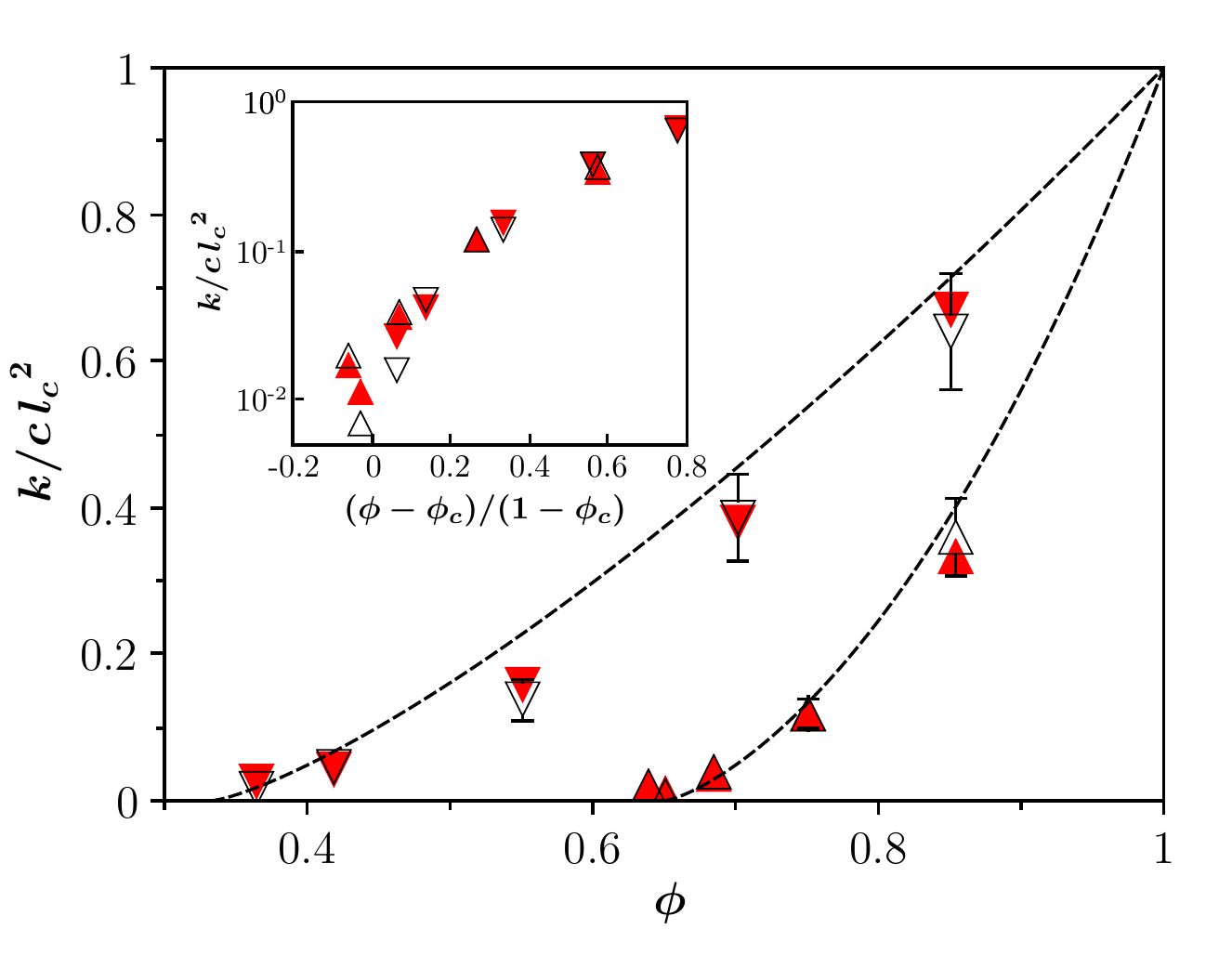}%
	\caption{\label{Fig:PermeabilityPhi}(Color online) Experimentally (open symbols) and numerically (closed symbols) measured permeability $k$ of ROMC (${\color{red}\blacktriangledown} , \triangledown$) and ROME structures (${\color{red}\blacktriangle} , \vartriangle$) as a function of the porosity $\phi$. All data points are normalised by $c\,{l_c}^2$. The dashed lines corresponds to the interpolation formula given in \cite{Xia:1988}. Inset: $k$ vs.\ the rescaled porosity, according to Eq.\ (\ref{Eq:CriticalArchie}). The increasing deviations between experimental and numerical data at small permeabilities are a result of discretization errors and numerical artifacts which increase close to $\phi_c$. Due to finite size effect some $\phi$ are shifted to negative values.}
\end{figure}
\begin{figure}
	\includegraphics[width=\figwidth]{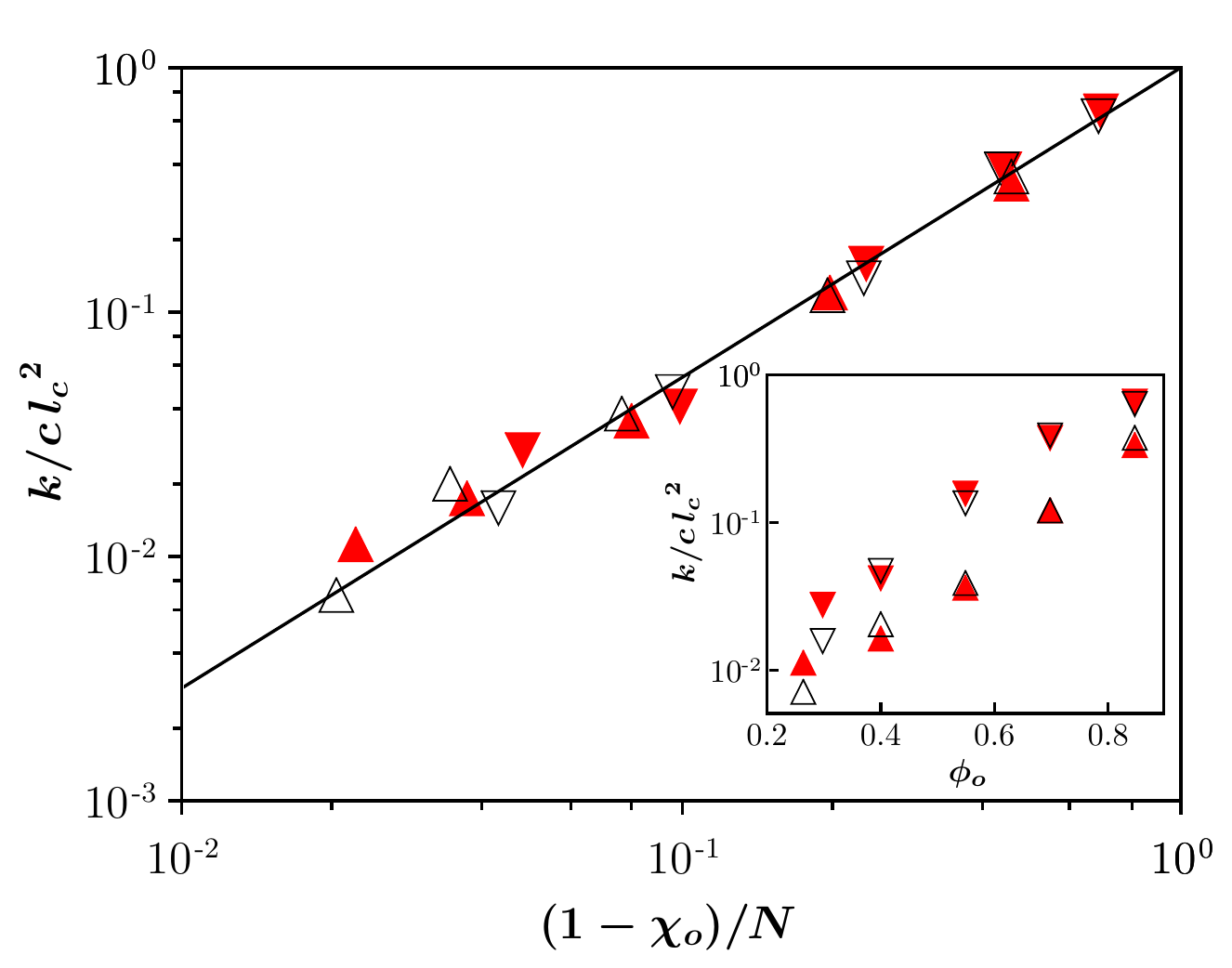}%
	\caption{\label{Fig:MeasurementTheory}(Color online) Experimentally (open symbols) and numerically (closed symbols) measured permeability of ROMC structures (${\color{red}\blacktriangledown} , \triangledown$) and ROME structures (${\color{red}\blacktriangle} , \vartriangle$) versus $(1-\chi_o)/N$. Error bars are smaller than the symbol size. The solid line is a fit of Eq.\ (\ref{Eq:EulerPermeability}) with $\alpha=1.27$. Inset: $k$ in dependence of the open porosity $\phi_o$.}
\end{figure}
\begin{figure}[t!]
	\includegraphics[width=\figwidth]{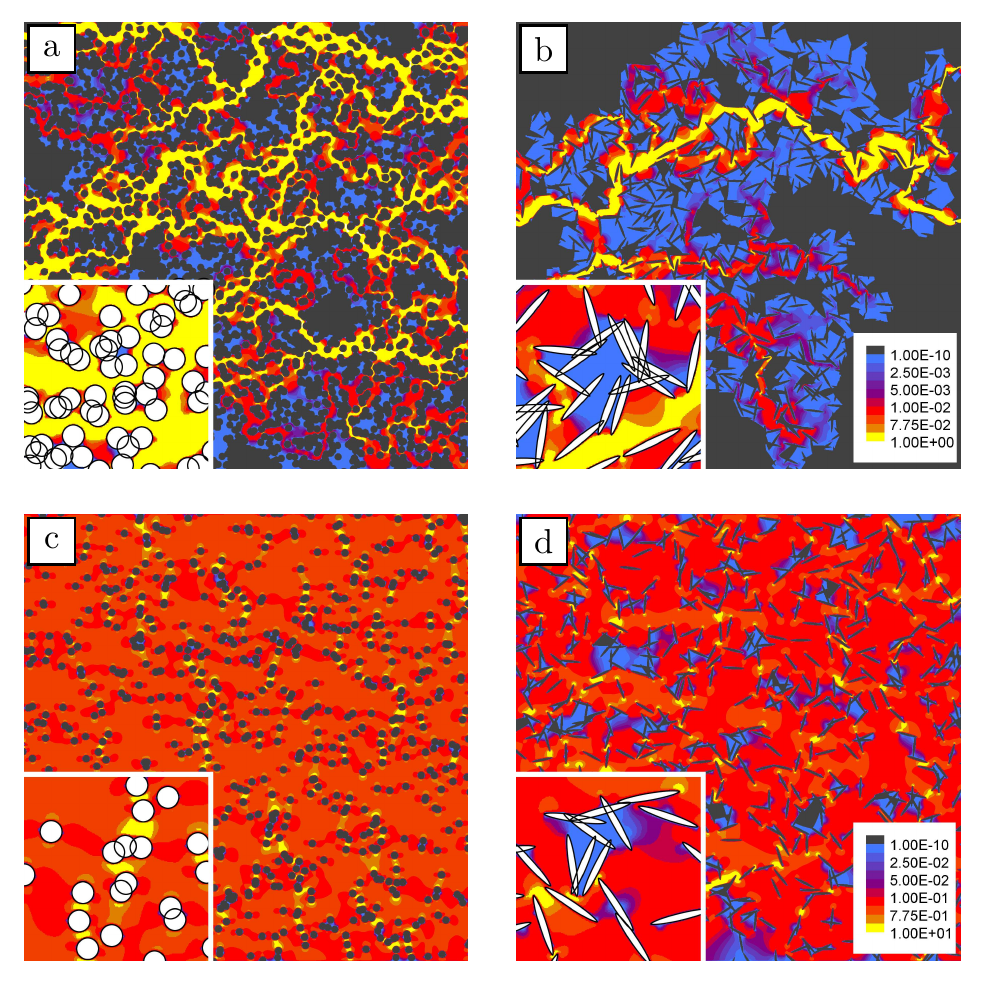}%
	\caption{\label{Fig:VelocityFields}(Color online) Contour plots of the simulated velocity field for two realizations of each process with equal open porosities $\phi_o=0.4$ (a,b) and $\phi_o=0.85.$ (c,d). The plots show the local average velocity magnitude. A logarithmic colorbar, that is identical for a,b and c,d respectively, is used for visualization. The unit is $\text{\textmu{}m}/\text{s}$. As illustrated in the insets elliptical grains form larger but less compact interconnected obstacles which leads to more tortuous streamlines and stagnant zones where the velocity goes to zero.}
\end{figure}

The understanding of liquid flow through porous materials is of huge practical importance in many fields of research such as soil and material science or biomedical applications (for a detailed review see \cite{Sahimi:1993}). In particular when the porous matrix has a non-regular structure, as typically encountered under natural conditions, it is an open question which structural information determines its permeability $k$, i.e., the rate at which a fluid can flow through the matrix at a given pressure difference $\Delta P$. Among various semi-empirical approaches aiming to relate the matrix properties to its capability for fluid flow, the Katz-Thomson model is one of the most established \cite{KatzThompson:1986}. It provides a link between $k$ and the conductivity $\sigma$ of the fluid phase within a porous material 
\begin{equation}
k=c\, {l_c}^2 \left(\frac{\sigma}{\sigma_0} \right),
\label{Eq:KatzThompson}
\end{equation}
where $c$ is a constant which depends on the local pore geometry, $l_c$ the critical pore diameter and $\sigma_0$ the fluid's bulk conductivity. A particular strength of this model is the identification of a characteristic length scale $l_c$ which is set by the largest diameter of a spherical particle which can freely pass the porous medium. Experimental and numerical studies have confirmed the applicability of Eq.\ (\ref{Eq:KatzThompson}) and in particular the dependence of $k$ on $l_c$ for different types of natural rocks \cite{Martys:1992, Arns:2005, Quispe:2005, Andersson:2011}. In general, it is not obvious how $\sigma/\sigma_0$ is linked to the structural properties of the porous matrix. The most successful approach is Archie's law
\begin{equation}
\frac{\sigma}{\sigma_0} = \left(\frac{\phi-\phi_c}{1-\phi_c}\right)^\mu ,
\label{Eq:CriticalArchie}
\end{equation}
where $\phi$ is the porosity, i.e., the volume fraction of the liquid phase, $\phi_c$ the corresponding value at the percolation threshold (the highest value for $\phi$ below which there is no sample-spanning liquid phase) and $\mu$ a critical exponent. Eq.\ (\ref{Eq:CriticalArchie}) has been extensively studied by numerical simulations for discrete and continuum percolation models. Most authors observe a universal value $\mu=1.3$ in two-dimensional systems \cite{Kirkpatrick:1973, Halperin:1985, Tobochnik:1990}, however, deviations from this value have also been found \cite{Sen:1985,Bunde:1986,Octavio:1988}. Despite its simplicity, the application of Archie's law to simulations and experiments and in particular to natural porous materials, is limited because (i) Eq.\ (\ref{Eq:CriticalArchie}) is only valid close to the percolation threshold and even more important by (ii) $\phi_c$ cannot be obtained from a single sample but is only defined when the underlying formation (diagenetic) process of the porous material is known. These limitations may explain the partially controversial results obtained for $\mu$ in the literature.\\
To overcome the above problems, a morphological reconstruction of porous materials based on Boolean models, i.e.\ randomly placed overlapping grains, has been proposed \cite{Arns:2003,Mecke:2005,Lehmann:2008}. For sedimentary rock samples, it has been demonstrated that their permeability and mechanical stiffness are very close to those of Boolean models for suitably chosen parameters \cite{Arns:2003}. These parameters can be determined from a specific set of morphological measures, so-called Minkowski functionals \cite{Schroeder:2011}, which allow to characterize the morphology of random geometries and have already been successfully employed for a wide range of problems, e.g.\ dewetting patterns of polymer films or the large-scale structure of the universe \cite{Mantz:2008,Mecke:1994}. So far, it is not clear whether a direct relation between the permeability and the Minkowski functionals exists. In addition to reducing the computational effort to obtain such quantities, a direct link between the permeability and morphological measures would allow for a better understanding of fluid flow through porous materials.
\\
In this Letter we study the permeability of quasi two-dimensional (q2D) porous structures of randomly placed overlapping monodisperse circles and ellipses. We performed measurements in microfluidic devices and lattice Boltzmann simulations which demonstrate that the permeability is determined by the Euler characteristic (an integral geometric measure related to the connectivity of the pore space) of the conducting phase. We introduce an expression for the permeability that -- in contrast to Eq.\ (\ref{Eq:CriticalArchie}) -- 
is independent of the percolation threshold and only depends on morphological properties of an individual sample. The expression is found to be in excellent agreement with our data over a wide range of porosities. 
\\
For our investigations we used structures created by Boolean models with two different types of grains: (i) randomly placed overlapping monodisperse circles (ROMC) with constant radius $r$ and (ii) randomly placed overlapping monodisperse ellipses (ROME) with constant aspect ratio $a/b=8$ with $a$ and $b$ the long and short semiaxis, respectively. Both, the position and (in case of the ellipses) the orientation of grains is random and uniformly distributed.
\\
In our experiments, microfluidic channels made of polydimethylsiloxane (PDMS) with dimensions $8-10\,\text{\textmu{}m}$, $3.5\,\text{mm}$, $10-11\,\text{mm}$ (height $h$, width, length) were created by soft lithography \cite{Whitesides:2001}.
The center of such straight channels contained q2D ROMC and ROME structures extending over a length $L= 3.5\,\text{mm}$. In our samples we set $r=30\,\text{\textmu{}m}$ and $a=84\,\text{\textmu{}m}$. \\
As illustrated in Fig.\ \ref{Fig:ExperimentStructures}a we apply a hydrostatic pressure in the range of $\Delta P=10-50\, \text{Pa}$ to the sample by connecting two reservoirs to the in- and outlet and varying the amount of injected liquid. In order to determine the flow properties through the structures, a diluted aqueous suspension of $1.3\,\text{\textmu{}m}$ polystyrene tracer particles is injected into the channel. From the mean particle velocity which is measured by digital video microscopy, we obtain the averaged fluid velocity $\bar v$ (for details see \cite{footnote1,Scholz:2012}). The permeability of the structures is then easily obtained by application of Darcy's law
\begin{equation}
q = -\frac{k}{\eta} \frac{\Delta P}{L} ,
\end{equation}
where $\eta$ is the viscosity of the fluid and $q=\bar{v}\cdot\phi$.\\
In addition to experimentally measured values of $k$, we determined the permeability by numerical simulations of the fluid flow through the structures using the massively parallel lattice Boltzmann application framework \textsc{wa}LB\textsc{erla} \cite{Koerner:2006, Luo:1998}. The experimental structures were modeled by a lattice of size $8 \times 4000 \times 4000$. 
In the simulations the pressure was set to $\Delta P=50\,\text{Pa}$, comparable to the experiments.
\\
\begin{table}
	\caption{\label{Tab:ROMC}Morphological properties of the studied realizations of the ROME and ROMC models. $D_c$ and $l_c$ are given in units of lattice sites.}
	\subfloat[ROMC][ROMC]{
		\begin{ruledtabular}
			\begin{tabular}{cccccc}
				$\phi$ & $\phi_o$ & N & $\chi_o$ & $D_c$ & $l_c$ \\
				\hline \\
				0.365 & 0.298 & 4592 & -220 & 2.421  & 2.421 \\
				0.418 & 0.401 & 3968 & -395 & 11.506  & 8 \\
				0.551 & 0.549 & 2704 & -635 & 21.984 & 8 \\
				0.701 & 0.700 & 1632 & -724 & 51.306 & 8 \\
				0.850 & 0.850 & 754  & -520 & 100.018 & 8 
			\end{tabular}
		\end{ruledtabular}}\\
		\subfloat[ROME][ROME]{
		\begin{ruledtabular}
			\begin{tabular}{cccccc}
				$\phi$ & $\phi_o$ & N & $\chi_o$ & $D_c$ & $l_c$ \\
				\hline \\
				0.651 & 0.266 & 2064 & -45  & 5.957  & 5.957 \\ 
				0.639 & 0.400 & 2176 & -80  & 6.245 & 6.245 \\
				0.684 & 0.549 & 1840 & -146 & 6.275 & 6.275 \\
				0.751 & 0.700 & 1387 & -275 & 41.254 & 8 \\
				0.854 & 0.850 & 771  & -352 & 67.429 & 8 
			\end{tabular}
		\end{ruledtabular}}
\end{table}	
We studied structures with five different porosities created according to the ROMC and ROME algorithm. The porosities were chosen to cover a range between $\phi\approx 0.85$ down to the corresponding percolation thresholds $\phi_c\simeq 0.32$ (ROMC) and $\phi_c\simeq 0.66$ (ROME) which are estimates from Monte Carlo simulations \cite{Quintanilla:2007, Yi:2002}. Each sample is also characterized by its \emph{open porosity} $\phi_o$ corresponding to the volume fraction in which fluid flow takes place (i.e.\ the conducting phase excluding regions where liquid is trapped in closed cavities, see Fig.\ \ref{Fig:ExperimentStructures}b) and the grain density $n=N/L^2$, where N is the total number of circular or elliptical grains and $L$ the linear system size. The different morphological quantities relevant throughout this paper are summarized in Tab.\ \ref{Tab:ROMC}. 
\\
For the q2D structures considered here, it is important to realize that the critical pore diameter introduced in Eq.\ (\ref{Eq:KatzThompson}) can be limited by the sample height $h$. When the two-dimensional critical pore diameter $D_c$, i.e. the size of the largest disk that freely penetrates the structure is larger than $h$, the latter defines the limiting hydrodynamic length scale, i.e.
$l_c = \text{min}(D_c,h)$.
The values of $D_c$ for our samples have been numerically computed from the Euclidean distance transform \cite{Hilpert:2001,Mickel:2008} of the structures and are also shown in Tab.\ \ref{Tab:ROMC}. Two typical structures and their corresponding critical pores (red arrows) are illustrated in Fig.\ \ref{Fig:ExperimentStructures} c,d.
\\
Fig.\ \ref{Fig:PermeabilityPhi} shows the experimentally (open symbols) and numerically determined (closed symbols) permeability (normalized by $c {l_c}^{2}$) vs the porosity. As expected, both sample types exhibit a decreasing permeability when the porosity is lowered towards its corresponding percolation threshold. Our results are consistent with effective medium theories for $\sigma/\sigma_0$ which interpolate between Eq.\ (\ref{Eq:CriticalArchie}) and the dilute limit \cite{Xia:1988, Tobochnik:1990} (dashed line). When rescaling the porosity according to Eq.\ (\ref{Eq:CriticalArchie}), all ROMC and ROME data collapse to a single curve demonstrating the validity of Archie's law (inset of Fig.\ \ref{Fig:PermeabilityPhi}) and that the transport properties can be described by a master curve even far away from $\phi_c$ (note, that individual data points deviate from this curve due to the finite system size). However, the biggest drawback of Eq.\ (\ref{Eq:CriticalArchie}) is the dependence on the percolation threshold which requires full knowledge of the diagenetic process.\\
In the following, we propose an alternative expression for the permeability which does not require knowledge about $\phi_c$ but only depends on the morphological properties obtained from the specific sample. As mentioned above, morphological properties can be quantified by the use of Minkowski functionals, defined by integral geometry. With respect to scalar additive functionals, it can be proven that only a finite number of functionals provide independent shape information \cite{Klein:1997}. In two dimensions, the Minkowski functionals are (i) the area, (ii) the perimeter between the solid and liquid phase and (iii) the Euler characteristic $\chi$ which is the difference between the number of connected components of each phase \cite{Mecke:2005}. We also distinguish $\chi$ and $\chi_o$, i.e.\ the Euler characteristic of the conducting, i.e. liquid phase, because $k$ is not affected by cavities inside solid clusters, which do not contribute to the flow. In Fig.\ \ref{Fig:ExperimentStructures}b we provide a simple geometrical interpretation of these morphological quantities.
\\
Motivated by the observation, valid in various stochastic processes, that the Euler characteristic vanishes near the percolation threshold \cite{Mecke:1991,Neher:2008}, we propose the following empirical expression for the permeability of a porous structure
\begin{equation}
	k=c\, {l_c}^2 \left(\frac{1-\chi_o}{N} \right)^\alpha .
	\label{Eq:EulerPermeability}
\end{equation}
For a q2D geometry as considered here, the local pore geometry parameter introduced in Eq.\ (\ref{Eq:KatzThompson}) is $c=1/12$ which corresponds to the geometrical confinement between two parallel plates \cite{Bruus:2007}. From the geometrical interpretation of $\chi_o$ given in Fig.\ \ref{Fig:ExperimentStructures}b it follows that $1-\chi_o$ (typically referred to as genus) is equal to the total number of solid clusters formed by isolated or overlapping grains. Consequently the factor $(1-\chi_o)/N$ in Eq.\ (\ref{Eq:EulerPermeability}) is the fraction of such clusters per grain within the porous material. The limiting cases of Eq.\ (\ref{Eq:CriticalArchie}) are correctly reproduced. For $\phi \rightarrow 1$ solid grains hardly overlap this leads to $(1-\chi_o)\rightarrow N$. In contrast, when $\phi \rightarrow \phi_c$, a single percolating solid cluster will form which leads to $(1-\chi_o)/N\rightarrow 1/N \rightarrow 0$. 
\\
Fig.\ \ref{Fig:MeasurementTheory} shows the permeabilities obtained from experiments (open symbols) and simulations (closed symbols) as a function of $(1-\chi_o)/N$. Independent of the shape of the grains forming the porous structure, the data points which cover more than two decades, are described by a single straight line with slope $\alpha$ in accordance with Eq.\ (\ref{Eq:EulerPermeability}). Best agreement is found for $\alpha=1.27\pm 0.09$ (solid line), close to the critical conductivity exponent $\mu=1.3$ (Eq.\ (\ref{Eq:CriticalArchie})). 
\\
The good agreement between our data and Eq.\ (\ref{Eq:EulerPermeability}) suggests that independent of the shape of single grains, it is their aggregation into solid clusters which dominates the permeability in porous materials. In contrast to isolated grains, where the liquid can easily flow around, extended clusters lead to tortuous streamlines and thus reduce the sample's permeability. As a result of the different overlapping probabilities, clusters form in ROMC and ROME structures at different $N$. This rationalizes why the fraction of clusters per grain enters Eq.\ (\ref{Eq:EulerPermeability}). The importance of extended clusters for the permeability is also supported by the microscopic flow fields obtained from our numerical simulations. In Fig.\ \ref{Fig:VelocityFields} we compare the absolute velocity in the horizontal plane averaged over the vertical direction for a ROMC and ROME structure with $\phi_o=0.4$ and 0.85, respectively. At the same open porosity, elliptical grains form larger less compact clusters than spherical grains (insets of Fig.\ \ref{Fig:VelocityFields}a,b and c,d). This leads to the formation of stagnant zones with vanishing fluid velocity in ROME structures and thus to a decreasing permeability. This also explains why $k$ of ROME structures is always smaller than that of ROMC structures with equal $\phi_o$ as shown in the inset of Fig.\ \ref{Fig:MeasurementTheory}. 
\\
We expect that Eq.\ (\ref{Eq:EulerPermeability}) should also hold in three dimensions and for other types of grains, in particular for any grains with convex shapes. For porous structures where the grain density is unknown or ill-defined (e.g.\ bio-networks) our expression could be applied by using a Boolean reconstruction. This method allows one to derive an appropriate Boolean model (e.g.\ polydisperse mixture of grains) with an effective grain density $\hat{n}$ from the full set of Minkowski functionals, so that Eq.\ (\ref{Eq:EulerPermeability}) can be used \cite{Arns:2009}. We hope that this work will stimulate further research to explore the range of porous materials where Eq.\ (\ref{Eq:EulerPermeability}) can be applied.
\\
In conclusion we have determined the permeability of porous q2D structures of overlapping circles and overlapping ellipses. We find quantitative agreement between measured and numerical data and with an empirical expression or $k$ based on the Euler characteristic of the structure. Our expression predicts $k$ over a wide range of porosities and does not require knowledge of the percolation threshold but is related to the fraction of clusters formed by overlapping grains.

\begin{acknowledgments}
We would like to thank Yujie Li and Jakob Mehl for inspiring discussions.
\end{acknowledgments}


\begin{thebibliography}{33}
\bibitem [{foo()}]{footnote1}%
{The fluid velocity $\bar v$ can be
  determined from the mean particle velocity $\bar u$. although in general
  $\bar u \neq \bar v$ it was shown that $\bar v=c\cdot\bar u$, where
  $c=\text{const}$, so that only a calibration in a reference channel is
  required.}%
\end{thebibliography}

%

\end{document}